\documentclass[12pt]{article}

\usepackage{amsmath}
\usepackage{amssymb}
\usepackage{amsfonts}
\usepackage{latexsym}
\usepackage{color}

\catcode `\@=11 \@addtoreset{equation}{section}

\catcode `\@=12

\newcommand{\be}{\begin{equation}}
\newcommand{\en}{\end{equation}}
\newcommand{\bea}{\begin{eqnarray}}
\newcommand{\ena}{\end{eqnarray}}
\newcommand{\beano}{\begin{eqnarray*}}
\newcommand{\enano}{\end{eqnarray*}}
\newcommand{\bee}{\begin{enumerate}}
\newcommand{\ene}{\end{enumerate}}

\newcommand{\mc}{\mathcal}

\newcommand{\D}{{\mc D}}

\newcommand{\E}{{\cal E}}

\newcommand{\1}{1 \!\! 1}

\newcommand{\Hil}{\mc H}

\catcode `\@=11 \@addtoreset{equation}{section}
\catcode `\@=12

\begin{document}

\thispagestyle{empty}

\vspace*{2cm}

\begin{center}
{\Large \bf Quons, coherent states and intertwining operators}   \vspace{2cm}\\

{\large F. Bagarello}\\
  Dipartimento di Metodi e Modelli Matematici,
Facolt\`a di Ingegneria,\\ Universit\`a di Palermo, I-90128  Palermo, Italy\\
e-mail: bagarell@unipa.it

\end{center}

\vspace*{2cm}

\begin{abstract}
\noindent We propose a differential representation for the  operators satisfying the q-mutation relation $BB^\dagger-q\,B^\dagger B=\1$ which generalizes a recent result by Eremin and Meldianov, and we discuss in detail this choice in the limit $q\rightarrow1$.  Further, we  build up non-linear  and Gazeau-Klauder coherent states associated to the free quonic hamiltonian $h_1=B^\dagger B$. Finally we construct almost isospectrals quonic hamiltonians adopting the results on intertwining operators recently proposed by the author.

\end{abstract}

\vspace{2cm}

%{\bf PACS Numbers}:  .......

\vfill

%\pagenumbering{roman}

\newpage

\section{Introduction}

In a recent paper, \cite{eremel}, the authors have proposed an explicit representation for the quons operators,  \cite{moh, andr} and references therein.
These are
defined essentially by their q-mutation relation \be [B,B^\dagger]_q:=B B^\dagger -q B^\dagger B
=\1, \qquad q\in [-1,1], \label{11} \en
between the creation and the annihilation operators $B^\dagger$
and $B$, which reduces to the canonical commutation relation for $q=1$ and to the canonical anti-commutation relation for $q=-1$. For  $q$ in
the interval $]-1,1[$, equation (\ref{11}) describes particles which are neither
bosons nor fermions. Other possible q-mutator relations have also been proposed along the years, but they will not be considered here. Our results generalize those in \cite{eremel}. More in detail we will find a differential expression
of the quon operators which reduce to the ones in \cite{eremel} under special situations. Moreover, we will construct the eigenstates of the hamiltonian $h_1=B^\dagger\,B$, and, using these states, we will also introduce two different families of coherent states, which will be compared with the coherent states existing in the literature, \cite{kar,baz}.

We also discuss some application of intertwining operators in this context, adopting the strategy proposed in \cite{bag1,baglast}, which produces, starting from $h_1$ and an operator $X$ such that $[h_1,X\,X^\dagger]=0$ and $X^\dagger\,X$ is invertible, a new hamiltonian $h_2$ whose eigensystem is related in a direct way to that of $h_1$.

\section{Generalities on quons}

Let $B$ and $B^\dagger$ satisfy the q-mutator in (\ref{11}) and let $\varphi_0$ be the vacuum of $B$: $B\varphi_0=0$.
In \cite{moh} it is proved that the eigenstates of $N_0=B^\dagger\,B$ are analogous to the bosonic ones, but for the normalization. More in details, putting
\be\varphi_n=\frac{1}{\beta_0\cdots\beta_{n-1}}\,{B^\dagger}^n\,\varphi_0=
\frac{1}{\beta_{n-1}}B^\dagger\varphi_{n-1},\qquad n\geq 1,\label{21}\en
we have $N_0\varphi_n=\epsilon_n\varphi_n$, with $\epsilon_0=0$, $\epsilon_1=1$ and $\epsilon_n=1+q+\cdots+q^{n-1}$ for $n\geq 1$. Also, the normalization is found to be $\beta_n^2=1+q+\cdots+q^n$, for all $n\geq0$. Hence $\epsilon_n=\beta_{n-1}^2$ for all $n\geq1$.  The set of the $\varphi_n$s  spans the Hilbert space $\Hil$ and they are mutually orhonormal: $<\varphi_{n},\varphi_{k}>=\delta_{n,k}$. Moreover, an useful consequence of (\ref{11}) is the following:
\be
B\varphi_{n+1}=\beta_n\varphi_{n},
\label{23}\en
for all $n\geq 0$. In \cite{eremel} the authors propose the following differential expressions for the operators $B$ and $B^\dagger$:
\be
B=\frac{\exp\{-2i\alpha x\}-\exp\{i\alpha\frac{d}{dx}\}\exp\{-i\alpha x\}}{-i\sqrt{1-\exp\{-2\alpha^2\}}},\quad B^\dagger=\frac{\exp\{2i\alpha x\}-\exp\{i\alpha x\}\exp\{i\alpha\frac{d}{dx}\}}{i\sqrt{1-\exp\{-2\alpha^2\}}},\label{24}\en
where $\alpha$ is related to the deformation parameter $q>0$ by $\alpha=\sqrt{-\frac{\log(q)}{2}}$. Hence $0<\alpha<\infty$. The ground state of $B$ is found to be
\be
\Psi_0(x)=\frac{1}{\pi^{1/4}}\exp\left(-\frac{x^2}{2}+\frac{3}{2}i\alpha x\right),
\label{26}\en
$B\Psi_0(x)=0$, and the excited eigenfunctions $\Psi_n(x)$ are also deduced, see \cite{eremel}. The authors  show that, in the limit $q\rightarrow1$, the standard bosonic wave-functions and operators are recovered. They also use $\Psi_n(x)$ to build up a family of coherent states.

In this section we want to show how more general differential representations of (\ref{11}) can indeed be found. We also will discuss why and when the representation in (\ref{24})-(\ref{26}) should be conveniently adopted. Other results on the same problem are contained in \cite{andr,spi} and references therein. We will comment on that at the end of this section.

The starting point is given by (\ref{24}), which we try to generalize in the following way:
\be
B=\alpha_1(x)+\exp\left\{i\gamma\frac{d}{dx}\right\}\alpha_2(x),\quad
B^\dagger=\overline{\alpha_1(x)}+\overline{\alpha_2(x)}\exp\left\{i\gamma\frac{d}{dx}\right\},\label{27}\en
where we are interested in finding the most general expression for $\alpha_j(x)$. Here $\gamma$ is assumed to be a real quantity, so that $e^{i\gamma\frac{d}{dx}}$ is a self-adjoint operator. Since $e^{i\gamma\frac{d}{dx}}$ acts on a generic (square-integrable) function $f(x)$ as $e^{i\gamma\frac{d}{dx}}f(x)=f(x+i\gamma)$, the action of $[B,B^\dagger]_q$ on $f(x)$ can be easily computed and gives
$
[B,B^\dagger]_qf(x)=a_0(x)f(x)+a_1(x)f(x+i\gamma)+a_2(x)f(x+2i\gamma),
$
where $$\left\{
\begin{array}{ll}a_0(x)=(1-q)|\alpha_1(x)|^2, \\ a_1(x)=\alpha_1(x)\overline{\alpha_2(x)}+\overline{\alpha_1(x-i\gamma)}\alpha_2(x+i\gamma)-q\left( \overline{\alpha_1(x)}\alpha_2(x+i\gamma)+\alpha_1(x+i\gamma)\overline{\alpha_2(x)}\right),\\
a_2(x)=\alpha_2(x+i\gamma)\overline{\alpha_2(x-i\gamma)}-q\,\overline{\alpha_2(x)}\alpha_2(x+2i\gamma).
\end{array}
\right.
$$
Hence $[B,B^\dagger]_q=\1$ if and only if $a_0(x)=1$ and $a_1(x)=a_2(x)=0$. It is easy to check that the one given in \cite{eremel}, see equation (\ref{24}), is not the most general solution of this problem. Let us roughly show how different solutions can be obtained. Condition $a_0(x)=1$ implies that $\alpha_1(x)=\frac{e^{i\Phi_1(x)}}{\sqrt{1-q}}$, for all possible real function $\Phi_1(x)$. Recall that, in agreement with all the existing literature on quons,  $q<1$. The case $q=1$ will be treated taking the limit $q\rightarrow1^-$. Condition $a_2(x)=0$ implies, in particular, that the ratio $\frac{\alpha_2(x+i\gamma)\overline{\alpha_2(x-i\gamma)}}{\overline{\alpha_2(x)}\alpha_2(x+2i\gamma)}$ must be independent of $x$ and equal to $q$. The simpler solution of this equation is $\alpha_2(x)=N_2e^{i\beta x}$, with $\beta$ satisfying $q=e^{2\beta\gamma}$. This can be explicitly checked. Of course, since $q<1$, we must also have $\beta\gamma<0$. Now we can replace what we have found so far in equation $a_1(x)=0$. This produces a condition on the unknown function $\Phi_1(x)$: $e^{i\Phi_1(x)}-e^{2\beta\gamma}e^{i\Phi_1(x+i\gamma)}=0$, which is solved taking, for instance, $\Phi_1(x)=2\beta x$.
Summarizing,  $B$ and $B^\dagger$ admit the following differential representation
\be
B=\frac{e^{2i\beta x}}{\sqrt{1-q}}+N_2\,e^{i\gamma\frac{d}{dx}}e^{i\beta x},\qquad B^\dagger=\frac{e^{-2i\beta x}}{\sqrt{1-q}}+N_2\,e^{-i\beta x}e^{i\gamma\frac{d}{dx}}.\label{29}\en
which reduces, but for an overall $i$, to the one in \cite{eremel} if $\beta=-\gamma=-\alpha$ and if $N_2=\frac{1}{\sqrt{1-q}}$.

As for the vacuum of the theory, we look for a solution of $B\varphi_0(x)=0$ as $\varphi_0(x)=K\,\exp\left\{b_1x^2+b_2x\right\}$, where $K$ is a normalization constant and $b_1, b_2$ should be fixed. It is easy to check that the most general solution of $B\varphi_0(x)=0$ of this form is the following function:
\be
\varphi_0(x)=\left(-\frac{\beta}{\pi\gamma}\right)^{1/4}\,\exp\left\{\frac{\beta}{2\gamma}\,
x^2-i\left(\frac{3}{2}\,\beta-\frac{1}{\gamma}\log(\nu_2)\right)\,x\right\}
\label{210}\en
Here $\nu_2$ is a real parameter which can assume any value larger than zero. Comparing these last two equations with (\ref{24}) and (\ref{26}) we see that our results are rather more general than those in \cite{eremel}. The two main evident differences are the presence of the ratio $\frac{\beta}{\gamma}$ in the normalization and as a coefficient of $x^2$ and the $\nu_2$-dependent phase  in $\varphi_0(x)$. This phase, of course, does not change the square modulus of the function. So one may wonder whether $\varphi_0(x)$ and $\Psi_0(x)$ are physically different or not, at least if $\beta=-\gamma$. Of course they are not, since the presence of  $\nu_2$  can play a relevant role in the computation of some matrix elements for particular operators. This is a well known fact in quantum mechanics: phases may matter!

\vspace{2mm}

Let us now define the following natural free hamiltonian for quons:
\be
h_1:=B^\dagger\,B=\frac{1}{1-q}\left(\1-\nu_2\,e^{-2\beta\gamma}\left(e^{i\beta x}+e^{-i\beta x}e^{\beta\gamma}\right)e^{i\gamma\frac{d}{dx}}+\nu_2^2e^{-2\beta\gamma}e^{2i\gamma\frac{d}{dx}}\right)
\label{211}\en
Its eigenstates are known to be, see (\ref{21}),
$
\varphi_{n+1}(x)=\frac{1}{\beta_n}\,B^\dagger \varphi_{n}(x)=\frac{1}{\beta_0\,\beta_1\cdots\beta_n}\,\left(B^\dagger\right)^{n+1} \varphi_{0}(x),
$
for all $n\geq0$. Since $h_1\equiv N_0$ we deduce that the eigenvalues associated to the $\varphi_{n}(x)$'s are the $\epsilon_n$'s introduced at the beginning of this section. In \cite{eremel} the authors give an explicit expression of the various $\varphi_{n}(x)$ in terms of $x$. Here we just give the first two excited states:
$
\varphi_{1}(x)=\frac{1}{\sqrt{1-q}}\,\left(e^{-2i\beta x}-e^{\beta \gamma}\right) \varphi_{0}(x),
$
and
$
\varphi_{2}(x)=\frac{1}{(1-q)\sqrt{1+q}}\,\left(e^{-4i\beta x}-e^{-2i\beta x}e^{\beta \gamma}\left(1+e^{2\beta\gamma}\right)+e^{2\beta\gamma}\right) \varphi_{0}(x),
$
while the others can be obtained recursively  as in (\ref{21}).

\subsection{Recovering the harmonic oscillator}

In \cite{eremel} the harmonic oscillator is recovered simply by taking the (formal) limit $q\rightarrow1^-$, which corresponds to the limit $\alpha\rightarrow0$. Here, due to the presence of more parameters, the situation is a bit different. We ask the following: how should our parameters be chosen for $B$ and $B^\dagger$ to collapse to the standard bosonic operators $b=\frac{1}{\sqrt{2}}\left(x+\frac{d}{dx}\right)$,  $b^\dagger=\frac{1}{\sqrt{2}}\left(x-\frac{d}{dx}\right)$, and for the eigenstates $\varphi_{n}(x)$ above to converge to the standard harmonic oscillator wave-functions?

It is convenient to start with the vacuum $\varphi_0(x)$ in (\ref{210}) which should tend, as $q\rightarrow1^-$, to the following gaussian: $f_0(x)=\frac{1}{\pi^{1/4}}\,e^{-x^2/2}$. This requirement forces $\beta$ and $\gamma$ to be related: we should have $\frac{\beta}{\gamma}\rightarrow -1$ in this limit. But, since $q\rightarrow1^-$ also implies that $\beta\gamma\rightarrow0^-$, the simplest choice is surely $\beta=-\gamma\rightarrow0$. With this choice we are forced to take $\nu_2=1$ in (\ref{210}), to avoid problems in the limit $\beta\rightarrow0$. Hence we have $\lim_{\beta\rightarrow0}\varphi_{0}(x)=f_0(x)$. These same assumptions allow us to conclude that $\lim_{\beta\rightarrow0}\varphi_{1}(x)=-if_1(x)$ and $\lim_{\beta\rightarrow0}\varphi_{2}(x)=-f_2(x)$, where $f_1(x)=\left(\frac{4}{\pi}\right)^{1/4}\,x\,e^{-x^2/2}$ and $f_2(x)=\left(\frac{1}{4\pi}\right)^{1/4}\,(2x^2-1)\,e^{-x^2/2}$ are the first two exited states of the harmonic oscillator.  Our claim is that, if $\beta=-\gamma$ and $\nu_2=1$, for all $n\geq0$ $\lim_{\beta\rightarrow0}\varphi_{n}(x)$ produces the n-th state of the harmonic oscillator $f_n(x)$ but, at most, for an overall power of $-i$.

An analogous analysis can be carried out for the creation operator $B^\dagger$ which becomes, if $\nu_2=1$ and $\beta=-\gamma$, $B^\dagger=\frac{1}{\sqrt{1-e^{-2\beta^2}}}\left(e^{-2i\beta x}-e^{-i\beta x}e^{-i\beta\frac{d}{dx}}\right)$. Therefore  $\lim_{\beta\rightarrow0^+} B^\dagger=-i\,b^\dagger$, where the limit must be understood in some weak operator sense. Hence $\lim_{\beta\rightarrow0^+} B=i\,b$. Again, but for an unessential overall $\pm i$, choosing $\nu_2=1$ and $\beta=-\gamma$ and sending $\beta$ to zero, we recover the standard bosonic operators.

This allows us to conclude that, even if the q-mutation relation can be represented in a rather general way by (\ref{29}), if we further require that the limit $q\rightarrow1^-$ returns the standard harmonic oscillator operators and wave-functions we are forced to impose extra conditions on the coefficients, and these conditions give back the definitions in \cite{eremel}. In other words, \cite{eremel} gives a representation of q-mutation relations with the extra requirement that the limit $q\rightarrow1^-$ returns the standard harmonic oscillator, which is a natural but not strictly necessary requirement. This clarify the reason for the choices in \cite{eremel}, where the operators were proposed with no justification at all.

\vspace{3mm}

Let us consider the following {\em physical} comment concerning the wave-functions $\varphi_n(x)$: it is clear that, if we consider for instance $|\varphi_0(x)|$ in (\ref{210}), this depends on $\frac{\beta}{\gamma}$ in such a way that, when the ratio $\left|\frac{\beta}{\gamma}\right|$ increases, then $|\varphi_0(x)|$ becomes more and more localized around $x=0$. On the other hand, if  $\left|\frac{\beta}{\gamma}\right|$ decreases, then $|\varphi_0(x)|$ delocalizes. Of course, these changes of $\frac{\beta}{\gamma}$ must preserve the normalization of the wave-function, and this is reflected by the value of $|\varphi_0(0)|$ due to the effect of $\left(-\frac{\beta}{\pi\gamma}\right)^{1/4}$.  Hence $\left|\frac{\beta}{\gamma}\right|$ behaves like a dilation parameter. This strongly reminds us what happens for ordinary wavelets. In both cases we have a dilation parameter. Here, however, the translation parameter is missing. Going back to the localization of the wave-functions, the same behavior is observed also for $\varphi_1(x)$ and $\varphi_2(x)$, which get more and more localized when $\left|\frac{\beta}{\gamma}\right|$ increases. For these functions (and, we believe, also for $\varphi_3(x)$, $\varphi_4(x)$ etc.), when $|\beta|$ is sufficiently different from $|\gamma|$, then the modulus of the functions are strictly concave, contrarily to what happens when $|\gamma|\simeq|\beta|$ and for the functions of the harmonic oscillator. The presence of $\left|\frac{\beta}{\gamma}\right|$ is important since the localization of the eigenstates of a given hamiltonian is usually crucial in many physical applications, for instance in the analysis of the ground state of a quantum gas of electrons, which is the concrete system used in the analysis of the quantum Hall effect, see \cite{bagant}. We should also mention that a dilation parameter appear in \cite{andr}, where a different possible differential representation of the (\ref{11}) is proposed. The surprising aspect is that, while in \cite{andr} a dilation operator appears already in the differential expression for the annihilation and creation operators, here we only use a translation operator in (\ref{27}), and the dilation parameter simply comes out. This suggests that maybe a {\em more complete} representation of (\ref{11}) would combine the two approaches. This is work in progress.

\section{Coherent states}

In this section we construct two different classes of coherent states associated to the quons and we prove that they satisfy a {\em minimal} number of properties required to any family of coherent states (CS). The problem of defining properly CS has been addressed in the literature by several authors and in several different ways, see \cite{alibag,cjt,gk,fhro} and references therein. These differences arise mainly because of the non-uniqueness of the definition of what a CS should be. To be more explicit, while some author defines them as eigenvectors of some sort of annihilation operators, \cite{cjt}, someone else appears more interested in getting a resolution of the identity, \cite{gk}. In some recent papers, \cite{baglast,alibag} and references therein, the authors have constructed different kind of vector CS associated to a general SUSY hamiltonians pair, which is still another kind of CS.

Before starting our analysis we should mention that in this section we will neglect the $x$-dependence in $\varphi_n$, since it will play no role: the results are infact representation-independent.

\subsection{Non-linear coherent states}

We begin our analysis by considering the following $z-$ depending vector, $z\in\Bbb{C}$:
\be
\Phi_z:=N^{-1/2}(|z|^2)\,\sum_{k=0}^\infty\,\frac{z^k}{\sqrt{\epsilon_k!}}\,\varphi_{k},
\label{31}\en
where $\{\epsilon_k\}$ is the sequence of eigenvalues of $h_1$ introduced at the beginning of Section II. Of course this sequence is such that $0=\epsilon_0<\epsilon_1<\epsilon_2<\cdots$. Hence $\epsilon_0!=1$ and we have defined, as usual, $\epsilon_k!=\epsilon_1\cdots\epsilon_k$, for $k\geq1$. The normalization $N$ turns out to depend only on $|z|^2$, rather than on $z$ itself. Indeed requiring that $<\Phi_z,\Phi_z>=1$ for all  $z$ we get that $
N(|z|^2)=\sum_{k=0}^\infty\frac{|z|^{2k}}{\epsilon_k!}$,
which converges if $|z|^2<\rho=\lim_{k,\infty}\epsilon_{k+1}=\frac{1}{1-q}$. Notice also that, for standard bosons, $\rho=\infty$. Hence $\Phi_z$ can be defined for all those $z\in\E$, where $\E=\{z\in\Bbb{C}\,:\,|z|^2<\rho\}$.

It is easy to check that $\Phi_z$ is an eigenstate of $B$ with eigenvalue $z$:
$B\Phi_z=z\Phi_z$.
This is a consequence of equation (\ref{23}) and of the relation between $\beta_n$ and $\epsilon_n$. Also, assuming that a measure $d\nu(r)$ exists such that the  moment problem $\int_0^\rho \,d\nu(r)\,r^{2k}=\epsilon_k!$ is solved,
 $k\geq0$, we can also deduce the following resolution of the identity:
$
\int_\E d\nu(z,\overline{z})\,|\Phi_z><\Phi_z|=\1$.
Here we are adopting the Dirac bra-ket notation and we have introduced the measure $d\nu(z,\overline{z})=N(|z|^2)d\nu(r)\frac{d\theta}{2\pi}$, $z=r\,e^{i\theta}$.  We will come back on the existence of $d\nu(r)$ below.

It is possible to show that these states saturate the Heisenberg uncertainty relation. Indeed, let us now introduce the following position and momentum-like operators associated to $B$ and $B^\dagger$:
$
X=\frac{1}{\sqrt{2}}\,\left(B+B^\dagger\right)$, $P=\frac{1}{\sqrt{2}\,i}\,\left(B-B^\dagger\right),
$
and the related quantities $(\Delta X)^2=<X^2>-<X>^2$, $(\Delta P)^2=<P^2>-<P>^2$. Here  $<A>$ is the mean value of the operator $A$ on the vector $\Phi_z$: $<X>=<\Phi_z,X\Phi_z>$, for instance. A straightforward computation shows that
$$
\Delta X=\Delta P=\sqrt{\frac{1}{2}(<BB^\dagger>-|z|^2)}\quad\Rightarrow\quad \Delta X \Delta P=\frac{1}{2}(<BB^\dagger>-|z|^2),
$$
which is equal to the mean value of $\frac{-i}{2}[X,P]$. Hence the Heisenberg uncertainty relation is indeed saturated, as stated above. Notice further that, in the bosonic limit, $<BB^\dagger>=1+|z|^2$, so that $\Delta X\, \Delta P=\frac{1}{2}$, as expected. For $q\neq1$, on the contrary, we get $<BB^\dagger>=1+q|z|^2$, so that $\Delta X \,\Delta P=\frac{1}{2}(1+|z|^2(q-1))$, which is surely less than $1/2$ if $q<1$. But, in order to have $\Delta X \,\Delta P\geq0$, we also must have $1+|z|^2(q-1)\geq0$, or $|z|^2\leq \frac{1}{1-q}$. This is always true because of the convergence condition we have found to ensure that $N(|z|^2)$, and therefore $\Phi_z$, exists.

\vspace{3mm}

It is interesting to compare this family of CS with the ones already existing in the literature and in particular with those in \cite{kar}. As a matter of fact, it is possible to check that these are exactly the same kind of CS, introduced following different strategies. The starting point is the following alternative expression for $\Phi_z$, which can be deduced using (\ref{21}):
\be
\Phi_z=N(|z|^2)^{-1/2}\sum_{k=0}^\infty\frac{(zB^\dagger)^k}{\epsilon_k!}\,\varphi_0=N(|z|^2)^{-1/2}N(zB^\dagger)\,\varphi_0.
\label{37}\en
In the literature, \cite{baz,eremel}, the function $N(s)$ is usually called the {\em q-exponential} since it reduces to the exponential function if $\epsilon_k=k$, as for standard bosons, and is a solution of a differential equation which contains the q-derivative rather than the usual one, \cite{baz,eremel}. We will not consider this aspect here. The scalar product between two such CS can now be conveniently expressed in terms of $N(s)$:
$
<\Phi_z,\Phi_w>=\frac{N(z\overline{w})}{\sqrt{N(|z|^2)\,N(|w|^2)}},
$
which shows that, as expected, these CS are not orthogonal. Using (\ref{37}) it is not hard to recognize that the CS defined in this section coincides with those defined in  \cite{eremel,kar,baz}, even if the way in which they have been introduced is different. We also want to stress that, because of this equivalence,  the existence of a measure satisfying the moment problem above is proved in \cite{kar}.

\subsection{Gazeau-Klauder like coherent states}

Here we will construct  the so-called Gazeau-Klauder CS, \cite{gk}, which are different from those discussed above and therefore  differ from the ones existing in the literature. These CS, labeled by two
parameters $J>0$ and $\gamma\in\Bbb{R}$, can be written in terms of the orthonormal
basis of  $h_1=h_1^\dagger$  as \be
|J,\gamma>=M(J)^{-1/2}\,\sum_{n=0}^\infty\,\frac{\,J^{n/2}\,e^{-i\epsilon_n\,\gamma}}{\sqrt{\rho_n}}\,\varphi_n.\label{41}\en
Requiring normalization of these states we find that $M(J)=\sum_{n=0}^\infty\,\frac{\,J^{n}\,}{\rho_n}$, which converges for $0\leq J<R$, $R=\lim_n \frac{\rho_{n+1}}{\rho_n}$ (which could be infinite).

These states satisfy the following properties:

\indent(1) the states $|J,\gamma>$ are {\em temporarily
stable}: $ e^{-ih_1t}\,|J,\gamma>=|J,\gamma+ t>$, $\forall
t\in\mathbb{R}$.
This is a general consequence of the eigenvalue equation $h_1\varphi_n=\epsilon_n\varphi_n$, $n\geq0$.

(2) if there exists a measure $d\rho(u)$, such
that $\int_0^R\,d\rho(u)\,u^n=\rho_n$ for all $n\geq 0$ then, introducing a two-dimensional
measure $d\nu(J,\gamma)=M(J)\,d\rho(J)\,d\nu(\gamma)$, with
$\int_{\mathbb{R}}\ldots
\,d\nu(\gamma)=\lim_{\Gamma\rightarrow\infty}\,\frac{1}{2\Gamma}\,
\int_{-\Gamma}^\Gamma\ldots\,d\gamma$, the following resolution of
the identity is satisfied: \be
\int_{C_R}\,d\nu(J,\gamma)\,|J,\gamma><J,\gamma|=\int_0^R\,M(J)\,\rho(J)\,dJ\,
\int_{\mathbb{R}}\,d\nu(\gamma)\,|J,\gamma><J,\gamma|=\1.
\label{43}\en
Here $C_R=\{(J,\gamma):\,J\in[0,R[, \gamma\in\Bbb{R}\}$.

At this stage we have not assumed any relation between  $\rho_n$ and the $\epsilon_n$'s. The only requirement on the $\rho_n$'s is that $R$ exists and is larger than zero (otherwise the CS are trivial!) and that the moment problem admits a solution. However, if we are also interested in the so-called {\em action identity}, \cite{gk},  $\rho_n$ must be related to $\epsilon_n$ in a rather stringent way. Indeed we find that if $\rho_k=\rho_0\,\epsilon_k!$ for all $k\geq1$, then

(3)  they also
satisfy the action identity: $
<J,\gamma|h_1|J,\gamma>=J.$

Because of the equation  $\rho_k=\rho_0\,\epsilon_k!$ we can now compute $R$ in terms of the eigenvalues of $h_1=B^\dagger\,B$, and again we get $R=\frac{1}{1-q}$.

\vspace{3mm}

The fact that $|J,\gamma>$ is different from $\Phi_z$ in (\ref{31}) is clearly shown by the fact that, for $q\neq1$, they are not eigenstates of $B$. This can be checked by a direct computation. However, even in this case, it is possible to introduce a $\gamma-$depending annihilation-like
operator $B_\gamma$ defined on $\varphi_n$ as follows:
\be
\!\!B_\gamma\,\varphi_n=\left\{
    \begin{array}{ll}
        0,\hspace{3.4cm}\mbox{ if } n=0,  \\
        \sqrt{\epsilon_n}\,e^{i(\epsilon_n-\epsilon_{n-1})\,\gamma}\varphi_{n-1}, \hspace{0.1cm} \mbox{ if } n>0\\
       \end{array}
        \right.=\left\{
    \begin{array}{ll}
        0,\hspace{2.8cm}\mbox{ if } n=0,  \\
        \sqrt{\epsilon_n}\,e^{i\,q^{n-1}\,\gamma}\varphi_{n-1}, \hspace{0.1cm} \mbox{ if } n>0,\\
       \end{array}
        \right.
\label{45}\en whose adjoint acts as
$B_\gamma^\dagger\,\varphi_n=\sqrt{\epsilon_{n+1}}\,e^{-i(\epsilon_{n+1}-\epsilon_{n})\,\gamma}\varphi_{n+1}=
\sqrt{\epsilon_{n+1}}\,e^{-i\,q^n\,\gamma}\varphi_{n+1}$. It is clear that for $q\neq1$ $B_\gamma\neq B$ in (\ref{23}). However, $B_\gamma$ is {\em very close} to $B$ since they coincide if $q=1$ or if $\gamma=0$. Moreover, for all possible $q$ and $\gamma$, $B_\gamma\,B_\gamma^\dagger=B\,B^\dagger$ and $B_\gamma^\dagger\,B_\gamma=B^\dagger\,B$.

With standard computations
we can also check that $ B_\gamma |J,\gamma>=\sqrt{J}\,
|J,\gamma>$. However, it should also be stressed that
$|J,\gamma>$ is not an eigenstate of $B_{\gamma'}$ if
$\gamma\neq\gamma'$.

\section{Intertwining operators}

In this section we discuss some results on quons in relations with intertwining operators, as given in \cite{baglast,bag1}. In particular we will see by considering some simple example that a purely algebraic vision of the system is much more convenient than the use of any concrete differential representation for the operators involved.

Let us first briefly recall our strategy, as given in \cite{baglast}. Let $h_1$ be a self-adjoint hamiltonian on the Hilbert space $\Hil_1$, $h_1=h_1^\dagger$, whose (not necessarily normalized) eigenvectors, $\varphi_n^{(1)}$, satisfy the following equation: $h_1\varphi_n^{(1)}=\epsilon_n\varphi_n^{(1)}$, $n\in\Bbb{N}$. Let $\Hil_2$ be a second Hilbert space, in general different from $\Hil_1$, and consider an operator $X:\Hil_2\rightarrow\Hil_1$, whose adjoint $X^\dagger$ maps $\Hil_1$ in $\Hil_2$. Let us further define $
N_1:=XX^\dagger$, $N_2:=X^\dagger X$.
It is clear that $N_j$ maps $\Hil_j$ into itself, for $j=1,2$. Suppose now that $X$ is such that $N_2$ is invertible in $\Hil_2$ and
$[N_1,h_1]=0$.
Of course this commutator should be considered in a {\em weak form} if $h_1$ or $N_1$ is unbounded: $<N_1f,h_1g>=<h_1f,N_1g>$, for $f,g$ in the domain of $N_1$ and $h_1$. Defining now
\be
h_2:=N_2^{-1}\left(X^\dagger\,h_1\,X\right),\qquad \varphi_n^{(2)}=X^\dagger\varphi_n^{(1)},
\label{53}\en
 the following conditions are satisfied:
 $h_2=h_2^\dagger$;
 $X^\dagger\left(X\,h_2-h_1\,X\right)=0$;
 if $\varphi_n^{(2)}\neq 0$  then $h_2\varphi_n^{(2)}=\epsilon_n\varphi_n^{(2)}$.
Furthermore, if $\epsilon_n$ is non degenerate, then $\varphi_n^{(1)}$  and $X^\dagger\varphi_n^{(1)}$ are eigenstates of $N_1$ and $N_2$ respectively with the same eigenvalue.

In the rest of this section we will restrict to a single Hilbert space: $\Hil=\Hil_1=\Hil_2$ and we consider, as first hamiltonian, the operator $h_1\equiv N_0=B^\dagger\,B$ introduced in Section II. The eigenstates are therefore $\varphi_n^{(1)}=\frac{1}{\beta_0\cdots\beta_{n-1}}\,{B^\dagger}^n\,\varphi_0^{(1)}=
\frac{1}{\beta_{n-1}}B^\dagger\varphi_{n-1}^{(1)}$,  $n\geq 1$,
with $B\varphi_0^{(1)}=0$. Hence $h_1\varphi_n^{(1)}=\epsilon_n\varphi_n^{(1)}$. Here $\epsilon_0=0$, $\epsilon_n=1+q+\cdots+q^{n-1}$ for $n\geq 1$ and $\beta_n^2=1+q+\cdots+q^n=\epsilon_{n+1}$, for all $n\geq0$.

\subsection{first example}
As a first example we consider the simplest possible situation, in which $X=B^\dagger$, $N_1=B^\dagger B$ and $N_2=BB^\dagger$.  Then obviously $[h_1,N_1]=0$. Moreover, since $N_2=BB^\dagger=\1+q\,B^\dagger B$, and since $B^\dagger B$ is a positive operator, we see that $N_2\geq\1$ in the sense of the operators if $q\geq0$. Hence $N_2^{-1}$ exists. Formula (\ref{53}) gives
$$
h_2:=N_2^{-1}\left(X^\dagger\,h_1\,X\right)=\left(BB^\dagger\right)^{-1}B(B^\dagger B)B^\dagger=B\,B^\dagger=q\,h_1+\1
$$
while
$$
\varphi_n^{(2)}=B\varphi_n^{(1)}= \left\{
\begin{array}{ll}
0\hspace{2.3cm} \mbox{ if } n=0  \\
\beta_{n-1}\varphi_{n-1}^{(1)}\qquad \mbox{ if } n\geq1\\
\end{array}
\right.
$$
Hence $h_2\varphi_n^{(2)}=(1+q\epsilon_{n-1})\varphi_n^{(2)}=\epsilon_{n}\varphi_n^{(2)}$. This means that, but for  $\epsilon_{0}=0$, all the other eigenvalues coincide, so that $\sigma(h_2)\subset\sigma(h_1)$.

It is worth to remark that, adopting an explicit differential procedure,  we would have been able to find the inverse of the operator
$$
N_2=B\,B^\dagger=\frac{1}{1-q}\left(\1-\nu_2\left(e^{i\beta x}+e^{-i\beta x}e^{\beta\gamma}\right)e^{i\gamma\frac{d}{dx}}+\nu_2^2e^{2i\gamma\frac{d}{dx}}\right),
$$
which looks quite a difficult problem to be solved, especially when compared with the algebraic approach proposed here. Notice that this example is just nothing but ordinary super-symmetric quantum mechanics, as discussed for instance in \cite{CKS}.

\subsection{generalizing this example}

The previous example can  be generalized to higher powers of $B^\dagger$: let $X=(B^\dagger)^l$, for all $l\geq 1$, and let us call $N_1^{(l)}=XX^\dagger=(B^\dagger)^l\,B^l$ and $N_2^{(l)}=X^\dagger\,X=B^l\,(B^\dagger)^l$. First of all we will prove that, for all $l\geq1$ and for $q\geq0$, $N_2^{(l)}$ admits inverse. Secondly we  check that, again for all $l\geq1$, $[h_1,N_1^{(l)}]=0$. Finally we  prove that the hamiltonian in (\ref{53}) can be written as
\be
h_2=q^l\,h_1+\sum_{k=0}^{l-1}q^k\1.
\label{55}\en
The proof of the  first statement uses induction on $l$: if $l=1$ then $N_2^{(1)}=B\,B^\dagger=\1+qB^\dagger\,B$ which, if $q\geq0$, is larger or equal to the identity operator $\1$ since $B^\dagger\,B\geq0$. Now, let us suppose that $N_2^{(l)}\geq\1$. Hence $$N_2^{(l+1)}=B^l\,B\,B^\dagger(B^\dagger)^l=B^l\,(\1+qB^\dagger B)\,(B^\dagger)^l=N_2^{(l)}+q\left(B^l B^\dagger\right)\left(B^l B^\dagger\right)^\dagger\geq N_2^{(l)}\geq\1. $$
The commutativity of $h_1$ and $N_1^{(l)}$ can be proved as follows:

first of all, using again induction we can check that, for all $l\geq1$,
\be
B^\dagger B^{l+1}=\frac{1}{q^l}\,B^l(B^\dagger B-(1+q+\cdots+q^{l-1})\1).
\label{55b}\en
This implies that, writing $N_1^{(l+1)}$ as $N_1^{(l+1)}=(B^\dagger)^{l+1}B^{l+1}=(B^\dagger)^l\,\frac{1}{q}(BB^\dagger-\1)\,B^l$, then
\be
N_1^{(l+1)}=\frac{1}{q^l}N_1^{(l)}\left(N_1^{(1)}-\sum_{k=0}^{l-1}q^k\1\right).
\label{56}\en
This suggests to use again the induction on $l$ to prove that $[h_1,N_1^{(l)}]=0$ for all $l\geq1$. Indeed we have, since $N_1^{(1)}=h_1$, $[h_1,N_1^{(1)}]=0$ trivially. Furthermore, assuming that $[h_1,N_1^{(l)}]=0$, equation (\ref{56}) and the first step of induction imply that $[h_1,N_1^{(l+1)}]=0$ as well.

Finally we want to recover expression (\ref{55}) for $h_2$. For this we need to compute equation (\ref{53}). Therefore, in principle, we should be able to find the inverse of the operator
$$
N_2^{(l)}=\frac{1}{(1-q)^l}\left(e^{2i\beta x}-\nu_2 e^{i\gamma\frac{d}{dx}}\,e^{i\beta x}\right)^l
\left(e^{-2i\beta x}-\nu_2 e^{-i\beta x}e^{i\gamma\frac{d}{dx}}\right)^l.
$$
This is a rather hard computation. However, it is easily seen that, as in the previous example, this can be avoided simply by using some consequences of the q-mutation relation $B\,B^\dagger-qB^\dagger\,B=\1$, and in particular the formula
\be
B\, (B^\dagger)^{l+1}= (B^\dagger)^{l}\left(q^l\,B^\dagger B+(1+q+\cdots+q^{l-1})\1\right),
\label{57}\en
which looks like the one in (\ref{55b}) but with $B$ and $B^\dagger$ exchanged. This can be proved again by induction on $l$. Formula (\ref{57})  allows us to write $\tilde h_2=X^\dagger h_1 X=B^l (B^\dagger\,B)(B^\dagger)^{l}$ as $\tilde h_2=B^l(B^\dagger)^{l} \left(q^l\,B^\dagger B+\sum_{k=0}^lq^k\1\right)$, so that $h_2=(N_2^{(l)})^{-1}\tilde h_2=q^l\,h_1+\sum_{k=0}^{l-1}q^k\1$, which is what we wanted to prove. It is evident that this approach is rather simpler than working directly in representation.

As for the eigenstates, if we put
$$
\varphi_n^{(2)}=B^l\varphi_n^{(1)}= \left\{
\begin{array}{ll}
0\hspace{4.4cm} \mbox{ if } n=0,1,\ldots,l-1  \\
\beta_{n-1}\beta_{n-2}\cdots\beta_{n-l}\varphi_{n-l}^{(1)}\qquad \mbox{ if } n\geq l,\\
\end{array}
\right.
$$
we can check that $h_2\varphi_n^{(2)}=\epsilon_n\varphi_n^{(2)}$ for all $n\geq l$.

\vspace{2mm}

{\bf Remarks:--} (1) We can obviously look at these results in a slightly different way: but for additive constant $\sum_{k=0}^{l-1}q^k$, which is crucial to ensure that the eigenvalue of $\varphi_n^{(2)}$ is exactly $\epsilon_n$, our intertwining operator $B^l$ produce, starting from $h_1=B^\dagger\,B$, a second hamiltonian $h=q^l \,B^\dagger\,B$ which has exactly the same eigenvectors as $h_1$, while the eigenvalues differ for an overall parameter $q^l$. This point of view could be interesting to produce pairs of non-isospectral hamiltonians, as in \cite{spi}. We plan to consider this aspect of the theory in a close future.

\vspace{2mm}

(2) We could think to generalize further the examples considered so far by taking $X=f(B^\dagger)$, where $f(z)$ is a analytic function which admits a power expansion in a region $\D\subset\Bbb{C}$.  This is because, as we have seen in Example 3, we are able to deal with any power of $B^\dagger$, so we may think that this procedure can be extended to such a function. However, we are immediately stopped since the requirement $[h_1,N_1]=0$ fails to be true, for general $f$. However, it is possible to check that
$$
\tilde h_2=X^\dagger h_1 X=f(B)\left(f(q\,B^\dagger)\,B^\dagger\,B+\frac{1}{1-q}\left(f(B^\dagger)-f(q\,B^\dagger)\right)\right),
$$
which is interesting since it appears in a natural way the q-derivative of $f(B^\dagger)$, \cite{baz}.

\vspace{2mm}

(3) Another similarity between ours and the results in \cite{andr} appears at this stage: it is sufficient to compare $h_1=B^\dagger\,B$ and $h_2$ in (\ref{55}) with the hamiltonians in \cite{andr} to check this similarity, which is due to the fact that the underlying methods used to derive them are quite close.

\vspace{3mm}

As already stated, the examples above show that finding $N_2^{-1}$ explicitly in the coordinate representation, for instance by looking for the Green's function of the operator, it is by far more difficult than using the algebraic results which follow directly from the q-mutation relations. In our opinion, working in representation should be left as the very last chance.

\section{Conclusions}

In this paper we have extended the differential representation of the q-mutation relations originally proposed in \cite{eremel} and we have shown that, with this extension, a natural dilation parameter appears which might be of some utility in physical applications. We have also considered in detail when and how the limit $q\rightarrow1^-$ returns the standard harmonic oscillator and we have shown that this is the requirement that makes our results to collapse with those in \cite{eremel}.

Then, starting from the quonic orthonormal functions, we have constructed two different families of CS, the so-called non-linear and the Gazeau-Klauder CS. The first class turns out to be already known in the literature, while the second ones are different and new, in this context.

Finally we have used quons to produce two examples of the general strategy of intertwining operators and spectra of hamiltonians, showing that it is much easier to adopt a purely algebraic point of view rather than any explicit representation of the operators involved.

\section*{Acknowledgements}

  The author acknowledges financial support by the Murst, within the  project {\em Problemi
Matematici Non Lineari di Propagazione e Stabilit\`a nei Modelli
del Continuo}, coordinated by Prof. T. Ruggeri.

\end{document}